\documentclass[10pt,conference,a4paper]{IEEEtran}
\usepackage[latin1]{inputenc}
\usepackage[cmex10]{amsmath}
\usepackage{amsfonts}
\usepackage{amssymb}
\usepackage{color}
\usepackage{tikz}

\newcommand{\R}{\mathbb{R}}

\title{A ``Piano Movers'' Problem Reformulated}

\author{David Wilson, James H. Davenport, Matthew England  \& Russell Bradford\\
Department of Computer Science, University of Bath, Bath, BA2 7AY, UK \\
E-mail: \texttt{\{D.J.Wilson, J.H.Davenport, M.England, R.J.Bradford\}@bath.ac.uk}}

\begin{document}

\maketitle

\begin{abstract}
It has long been known that cylindrical algebraic decompositions (CADs) can in theory be used for robot motion planning.  However, in practice even the simplest examples can be too complicated to tackle.  We consider in detail a ``Piano Mover's Problem'' which considers moving an infinitesimally thin piano (or ladder) through a right-angled corridor.

Producing a CAD for the original formulation of this problem is still infeasible after 25 years of improvements in both CAD theory and computer hardware.  We review some alternative formulations in the literature which use differing levels of geometric analysis before input to a CAD algorithm.  Simpler formulations allow CAD to easily address the question of the existence of a path.  We provide a new formulation for which both a CAD can be constructed and from which an actual path could be determined if one exists, and analyse the CADs produced using this approach for variations of the problem.   

This emphasises the importance of the precise formulation of such problems for CAD.  We analyse the formulations and their CADs considering a variety of heuristics and general criteria, leading to conclusions about tackling other problems of this form.
\end{abstract}

\section{Introduction}
\label{sec:Intro}

\subsection{A ``Piano Movers'' problem}
\label{subsec:pianomovers}

In \cite{SS83I} the authors describe a \emph{``Piano Movers'' Problem} as follows: ``given a body $B$ and a region bounded by a collection of walls, either find a continuous motion connecting two given positions and orientations of $B$ during which $B$ avoids collisions with the walls, or else establish that no such motion exists.''  Such problems commonly arise in robotics.

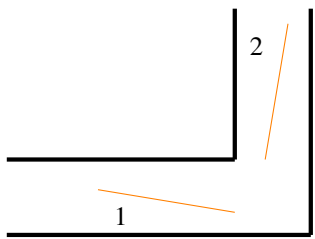
\begin{figure}[ht]
\centering
\begin{tikzpicture}
\draw[ultra thick] (0,0)--(-4,0);
\draw[ultra thick] (0,0)--(0,3);
\draw[ultra thick] (-1,1)--(-4,1);
\draw[ultra thick] (-1,1)--(-1,3);

\draw[orange] (-1.0,0.3)--(-2.8,0.6);
\draw[orange] (-0.6,1.0)--(-0.3,2.8);

\node [below] at (-2.5,0.5) {1};
\node [left] at (-0.5,2.5) {2};

\end{tikzpicture}
\caption{The piano movers problem considered in %\cite{Davenport:1986tf} 
\cite{Davenport86}}
\label{fig:originalDav}
\end{figure}

A simple example from %\cite{Davenport:1986tf} 
\cite{Davenport86} is the problem of moving a ladder of length 3 through a right-angled corridor of width 1 (moving from position 1 to position 2 in Figure \ref{fig:originalDav}).  
A simple analysis shows there is no solution to this particular problem, and that it would only be possible to traverse the corridor with a ladder of length less than $\sqrt{8}$.  We are interested in how this and similar piano movers problems may be decided automatically, with paths calculated when a solution exists.  

In %\cite{Schwartz:1983uk}
\cite{SS83II} the authors proposed a generic approach to piano movers problems in which the problem is described using polynomial algebra and then solved using the cylindrical algebraic decomposition (CAD) algorithm.  However, for even very simple examples this approach can be computationally infeasible.  In %\cite{Davenport:1986tf} 
\cite{Davenport86} the author applied the approach of %\cite{Schwartz:1983uk}
\cite{SS83II} to the simple problem of the ladder just described, demonstrating the scale of the computations that would be required.  Despite 25 years of improvements in both CAD theory and computer hardware, producing a CAD for the algebraic formulation given in %\cite{Davenport:1986tf} 
\cite{Davenport86} is still infeasible.

In Section \ref{sec:NewForm} we provide a new formulation for which a CAD has been produced and from which path could be deduced.  First we complete the introduction with a reminder of the theory of CAD, details of the original formulation and a summary of other formulations found in the literature.  Some of these can solve the existential question of whether a path exists very quickly, but they cannot then give the actual path that would be required by a robot, which the formulation presented in Section \ref{sec:NewForm} can.  In Section \ref{sec:Generalise} we consider generalisations of the problem and how the some of the formulations could be adapted while in Section \ref{sec:NewTech} we consider further adaptations to CAD technology for use with piano movers problems.  Finally we give our conclusions in Section~\ref{sec:Conclusion}.

\subsection{Cylindrical algebraic decomposition}
\label{subsec:cad}

A cylindrical algebraic decomposition (CAD) is a partition of $\R^n$ into cells, constructed with respect to an input, usually either polynomials or formulae, in $n$ ordered variables. Each cell is described by a semi-algebraic set (a finite sequence of polynomial equations and inequalities) and the cells are cylindrically arranged (meaning the projection of any two cells on the first $k$ coordinates is either equal or disjoint).  

A CAD is sign-invariant if the input polynomials have constant sign on each cell.  Such a CAD allows for the solution of many problems defined by the polynomials.  Collins provided the definition and first algorithm \cite{ACM84I}, motivated as a tool for quantifier elimination in real closed fields.  Other applications range from robot motion planning to algebraic simplification technology \cite{BD02, DBEW12}.

Collins' algorithm has two phases.  The first, \textit{projection}, applies a projection operator repeatedly to a set of polynomials, each time producing another set in one fewer variables.  Together these contain the {\em projection polynomials}.  The second phase, \textit{lifting}, then builds the CAD incrementally from these polynomials.  First $\R$ is decomposed into cells which are points and intervals corresponding to the real roots of the univariate polynomials.  Then $\R^2$ is decomposed by repeating the process over each cell using the bivariate polynomials at a sample point of the cell.  The output for each cell consists of {\em sections} of polynomials (where a polynomial vanishes) and {\em sectors} (the regions between these). Together these form the  {\em stack} over the cell, and taking the union of these stacks gives the CAD of $\R^2$.  This process is repeated until a CAD of $\R^n$ is produced.  The projection operator must be chosen in order to conclude that the CAD of $\R^n$ produced in this way is sign-invariant.
%To conclude that the CAD of $\R^n$ produced using sample points in this way ch that we need the key concept of delineability.  A polynomial is {\em delineable} in a cell if the portion of its zero set in the cell consists of disjoint sections.  A set of polynomials are {\em delineable} in a cell if each is delineable and further that the sections of different polynomials in the cell are either identical or disjoint.  A projection operator is valid for use in the algorithms if over each cell of a sign-invariant CAD for projection polynomials in $r$ variables, the polynomials in $r+1$ variables are delineable.

We note that CADs can depend heavily on the ordering of the variables.  In \cite{BD07} a problem was described which led to a cell count doubly exponential in the number of variables for one ordering, but constant in another.  Heuristics to help pick the variable ordering are developed in \cite{DSS04, BDEW13}.

Since Collins published the original algorithm there has been much research into improvements with a summary of developments over the first twenty years given by \cite{Collins98}.  Important advances include: the definition of finer projection operators to use in the first phase \cite{McCallum88}; the introduction of Partial CAD to make use of the quantified structure of a formula when lifting \cite{CH91}; the use of equational constraints to reduce the number of projection polynomials required \cite{McCallum99}; the use of truth-table-invariant CADs (TTICADs) to apply equational constraint techniques more widely \cite{BDEMW13}; and an alternative approach to projection and lifting where the problem is solved in complex space and then refined to a CAD of real space \cite{CMXY09}.

\subsection{Original formulation of the problem}
\label{subsec:JHD}

In %\cite{Davenport:1986tf} 
\cite{Davenport86} Davenport considered building a CAD to solve the problem of moving a ladder of length 3 through a right-angled corridor of width 1 (as in Figure \ref{fig:originalDav}). Denoting the endpoints of the ladder as $(x,y)$ and $(w,z)$ and assuming the outer corner of the corridor is the origin, the formulation provided was
%\begin{multline}
%\label{eq:originalpiano}
%\Big[ \big[(x - w)^2 + (y - z)^2 - 9 = 0 \big] \wedge
%\big[[y z \geq 0] \ \vee \\
%[x(y - z)^2 + y (w - x)(y - z) \geq 0] \big] \wedge
%\big[[(y -1)(z - 1) \geq 0] \\
%\vee [(x + 1)(y - z)^2 + (y - 1)(w - x)(y - z) \geq 0]\big] \wedge
%\big[[x w \geq 0] \\
%\vee \ [y (x - w)^2 + x(z - y)(x - w) \geq 0]\big] \wedge
%\big[[(x + 1)(w + 1) \geq 0]\\
%\vee [(y - 1)(x - w)^2 + (x + 1)(z - y)(x - w) \geq 0]\big] \ \Big].
%\end{multline}
\begin{align}
&\quad \big[(x - w)^2 + (y - z)^2 - 9 = 0 \big] 
\nonumber \\
&\wedge \big[[y z \geq 0] \vee [x(y - z)^2 + y (w - x)(y - z) \geq 0] \big] 
\nonumber \\
&\wedge \big[[(y -1)(z - 1) \geq 0] \nonumber \\
&\quad \vee [(x + 1)(y - z)^2 + (y - 1)(w - x)(y - z) \geq 0]\big] 
\nonumber \\
&\wedge \big[[x w \geq 0] \vee \ [y (x - w)^2 + x(z - y)(x - w) \geq 0]\big] 
\nonumber \\
&\wedge \big[[(x + 1)(w + 1) \geq 0] \nonumber \\
&\quad \vee [(y - 1)(x - w)^2 + (x + 1)(z - y)(x - w) \geq 0]\big]. 
\label{eq:originalpiano}
\end{align}

The first equation in \eqref{eq:originalpiano} describes the length of the ladder, and the remaining inequalities describe the valid positions, ensuring the ladder does not intersect any of the four walls.  
%In %\cite{Davenport:1986tf} 
%\cite{Davenport86}, \eqref{eq:originalpiano} was prefaced with existential quantifiers on $w$ and $z$ (but as the technology in %\cite{Collins:1991vz}
%\cite{CH91} was not known this had little effect on the discussion of the problem given there).

In %\cite{Davenport:1986tf} 
\cite{Davenport86} the author completed the projection phase of Collin's CAD algorithm, finding over 250 distinct univariate projection factors with total degree as high as 26.  The technology available for the paper did not allow for the simultaneous root isolation of these.  With current hardware\footnote{Experiments in this paper were run on a Linux desktop with a 3.1Ghz Intel processor and 8.0Gb total memory} 
and software incorporating the latest CAD theory ({\sc Qepcad-B} 1.69 \cite{Brown03b} and {\sc Maple} 16 \cite{CMXY09}) it still remains outside the realm of computation to complete the construction of the CAD. % Maple 16 doesn't do lifting as such

\subsection{Other approaches}
\label{subsec:other}

In robotics, piano mover's problems would typically be tackled using numerical methods to produce paths efficiently at the expense of the possibility of rounding errors.  We are concerned with the development of  symbolic approach and so do not examine numerical methods in this paper.

In %\cite{Schwartz:1983uk}
\cite{SS83I} the authors of \cite{SS83II} proposed a separate approach for the piano movers problems restricted to the plane, which did not make use of CAD.  This algorithm will typically run more efficiently than the CAD based approach but does not generalise to higher dimensional problems.

\begin{figure}[ht]
\centering
\begin{tikzpicture}
\draw[ultra thick] (0,0)--(-4,0);
\draw[ultra thick] (0,0)--(0,2);
\draw[ultra thick] (-1,1)--(-4,1);
\draw[ultra thick] (-1,1)--(-1,2);

\draw[orange] (-3.5,0)--(0,1.4);

\end{tikzpicture}
\caption{A configuration of a ladder in which the endpoints are in opposite branches of the corridor.
%and all four walls are intersected.  
}
\label{fig:Marchand}
\end{figure}
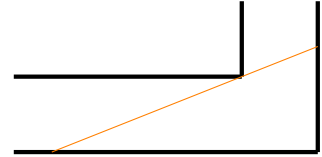

Tackling the problem with CAD has been revisited several times in the literature.  In %\cite{Marchand:1989aa}
\cite{Marchand89}, the author discussed the problem suggesting the question of traversing the corridor was equivalent to the question as to whether there is a position of the ladder for which both extremities are in the two branches of the corridor.  The verbal description of this reformulation seems misleading since a ladder could be positioned as such while still being unable to turn the corner as in Figure \ref{fig:Marchand}.

Later in %\cite{Marchand:1989aa}
\cite{Marchand89} the author reports on another reformulation using only one endpoint and the tangent of the half-angle between the $x$-axis and the ladder, reporting that a CAD can be produced for this using Collins' algorithm sufficient to conclude that the problem has no solution.  Since no details of the algebraic formulation were provided we are unable to verify this or analyse this formulation further.  

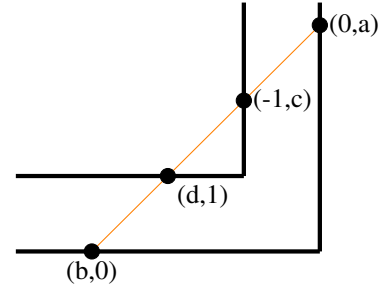
\begin{figure}[ht]
\centering
\begin{tikzpicture}
\draw[ultra thick] (0,0)--(-4,0);
\draw[ultra thick] (0,0)--(0,3.3);
\draw[ultra thick] (-1,1)--(-4,1);
\draw[ultra thick] (-1,1)--(-1,3.3);

\draw[orange] (-3,0)--(0,3);

\filldraw[fill=black, draw=black] (-3,0) circle [radius=0.1];
\node [below] at (-3,0) {(b,0)};
\filldraw[fill=black, draw=black] (0,3) circle [radius=0.1];
\node [right] at (0,3) {(0,a)};
\filldraw[fill=black, draw=black] (-2,1) circle [radius=0.1];
\node [below right] at (-2,1) {(d,1)};
\filldraw[fill=black, draw=black] (-1,2) circle [radius=0.1];
\node [right] at (-1,2) {(-1,c)};

\end{tikzpicture}
\caption{A configuration of a ladder in which all four walls are intersected.}
\label{fig:Wang}
\end{figure}

In %\cite{Wang:1996aa}
\cite{Wang96}, Wang uses ``simple reasoning'' to deduce that the ladder cannot  traverse the corridor if and only if it intersects all four walls simultaneously.
From this deduction the problem can be reformulated as follows:  Let $a,b,c,d$ be coordinates defining the intersection points as in Figure \ref{fig:Wang} and $r$ be the length of the ladder).  Then there is no solution if 
%\begin{multline}
%(\exists a)(\exists b)(\exists c)(\exists d)[a^2 + b^2 = r^2 \land a \leq 0 \land b \geq 0 \land c \leq -1 \\  \land d \geq 1 \land bc + a (1 - b) = 0 \land b + a (b - d) = 0].
%\end{multline}
\begin{align}
&(\exists a)(\exists b)(\exists c)(\exists d)[a^2 + b^2 = r^2 \land r>0 \nonumber \\
&\quad \land a \geq 0 \land b < 0 \land c \geq 1 \land d <-1 \nonumber \\
&\quad \land c-(1+b)(c-a)=0 \land d-(1-a)(d-b)=0]. \label{eq:Wang}
\end{align}
Due to its simplicity and the small number of free variables (only $r$ is unquantified) {\sc Qepcad} can almost instantly deduce that the maximal length of the ladder is $\sqrt{8}$, using a CAD of 19 cells.
%only 8.276 seconds and 35 cells? 
When considering the same problem in \cite{Wang91} Wang noted that if the ladder intersected the outer walls and one of the inner walls then it must also intersect the other.  Hence equation (\ref{eq:Wang}) could be simplified further by removing the final conditions on lines 2 and 3.  This further \emph{topological reasoning} actually makes no difference here ({\sc Qepcad}'s timings and cell counts are unchanged) but could be powerful for other problems.

In \cite{McCallum97} McCallum approaches path-finding by considering transformations of objects by a translation $(x,y)$ and a rotation $\theta$. This produces a formulation of the ladder problem involving 21 equations and inequalities in a comparatively complicated boolean formula. Appealing to equational constraints and partial CAD techniques McCallum constructs a four-dimensional CAD of 16,138 cells in 429 seconds.

In %\cite{Yang:2006aa}
\cite{YZ06} Yang and Zeng considered the problem in the case of a rectangular piano instead of a ladder and used geometric analysis to achieve a simple condition for the problem to have a solution. They parametrize the problem according to the position of a corner and the angle the rectangle makes with the horizontal axis.  Through some highly non-trivial analysis they obtain a condition on a polynomial which, if true, implies the existence of a valid route. Applying their techniques to the case of the ladder of length $L$ we see that the existence of a valid route is equivalent to the truth of %the following Tarski formula:
\begin{equation}
(\forall x) \, 4x^8 - 4(L-3)x^6-2(3L-6)x^4 - 2(L-3)x^2+1 > 0.
\end{equation}
It takes {\sc Qepcad} just 1.936 seconds (mostly initialization time) and 5 cells to return:
%\begin{equation}
$L^2-8<0  \lor L < 0.$
%\end{equation}

The approaches of %\cite{Wang:1996aa, Yang:2006aa}
\cite{Wang96} and \cite{YZ06} are highly efficient but limited.  They require, not insignificant, geometric deductions before presentation to CAD, and inform you only whether the ladder can or cannot pass through the corridor, revealing no information about possible paths. 
It would make sense to therefore use these sort of approaches as an initial test for a problem before constructing an inevitably far more-complicated CAD sufficient for planning routes.

Also, these reformulations give descriptions in the \emph{real space}, meaning they describe the geometry of the plane in which the ladder exists.  This is opposed to \cite{SS83II,Davenport86} and the new formulation in Section \ref{sec:NewForm} which describe the geometry in a four-dimensional \emph{configuration space}, specifically coordinates of the endpoints that fix the ladder within the plane.  This distinction is important since the former allows us to analyse whether a ladder can move through the corridor, but cannot provide the explicit path for it to do so.  It can be said that \cite{McCallum97} also works within a configuration space, however a non-trivial one where positions are encoded by transformations. 

By not considering the whole configuration space in their formulations, Wang and Yang--Zeng also cannot consider whether the ladder is able to rotate within the corridor to exit in the opposite orientation (an important point for the generalisations of the problem discussed in Section \ref{subsec:angled}).

\section{New formulation of the problem}
\label{sec:NewForm}

We consider the problem in configuration space, but from a different perspective than \cite{Davenport86}. First we give a formula describing all possible invalid regions, then take its negation as a description of the valid regions. As in \eqref{eq:originalpiano} we denote the endpoints of the ladder by $(x,y)$ and $(w,z)$. 

\subsection{Describing the invalid regions}

\begin{figure}[ht]
\centering
\begin{tikzpicture}
\draw[ultra thick] (0,0)--(-4,0);
\draw[ultra thick] (0,0)--(0,3.5);
\draw[ultra thick] (-1,1)--(-4,1);
\draw[ultra thick] (-1,1)--(-1,3.5);
\draw[orange] (-3.5,8/3)--(-1.5,7/3);
\draw[orange] (1/3,1.5)--(2/3,3.5);
\draw[orange] (-0.5,-1/3)--(-2.5,-2/3);
\draw[orange] (-2,0.5)--(-0.5,2);
\node [above] at (-2,2.5) {A};
\node [right] at (0.5,2.5) {B};
\node [below] at (-1.5,-0.5) {C};
\node [right] at (-1,1.5) {D};

\end{tikzpicture}
\\\vskip-10pt
\caption{Four canonical invalid positions of the ladder.  Note from the algebraic descriptions that for positions A--C only one end need be outside the corridor.}
\label{fig:ladderhallway}
\end{figure}
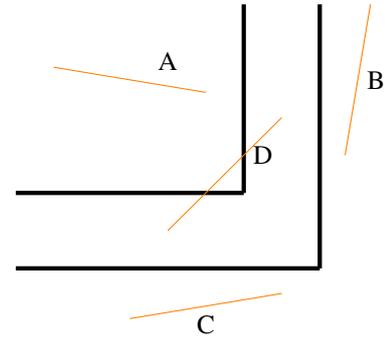

We describe four canonical invalid configurations for the ladder.  Each is identified with an equivalent Tarski formula and examples of each are given in Figure \ref{fig:ladderhallway}.
\begin{description}
  \item[A] $x < -1 \land y > 1$ or $w < -1 \land z > 1$: this describes any collision with the `inside' walls along with the ladder being on the other side of these.
  \item[B] $x > 0$ or $w > 0$: this describes any collision with the rightmost wall along with the ladder being on the other side.
  \item[C] $y < 0$ or $z < 0$: this describes any collision with the bottommost wall along with the ladder being on the other side.
  \item[D] $(\exists t) [ 0 < t \land t < 1 \land x + t(w-x) < -1 \land y + t(z-y)>1]$: this ensures no inner point of the ladder lies in the invalid top-left region. 
\end{description}

We can hence characterise the invalid regions with:
\begin{align}
&\quad [ x < -1 \land y > 1 ] \lor \left[ w < -1 \land z > 1 \right] 
\lor [x > 0] 
\nonumber \\ 
&\lor [w > 0] \lor  [y < 0] \lor [z < 0] 
\lor (\exists t) \big[ 0 < t \land t < 1 
\nonumber \\
&\quad \quad \land x + t(w-x) < -1 \land y + t(z-y)>1  \big].
\label{eq:nonprenexinvalid}
\end{align}
%which (as $t$ is not involved in any formulae outside the quantified one) can be rewritten in prenex form as:
%\begin{multline}\label{eq:prenexinvalid}
%(\exists t) \Big[ \left[ x < -1 \land y > 1 \right] \lor \left[ w < -1 \land z > 1 \right] \lor [x > 0] \lor [w > 0] \lor  [y < 0] \\ \lor   [z < 0] \lor \left[ 0 < t \land t < 1 \land x + t(w-x) < -1 \land y + t(z-y)>1  \right] \Big].
%\end{multline}
%
This formula contains the ``new'' variable $t$ used to represent any point on the ladder.  We can use {\sc Qepcad} to eliminate $t$ from \eqref{eq:nonprenexinvalid} in just over 2 seconds, %(including initialisation),
constructing 681 cells and returning the equivalent quantifier-free formula:
%\begin{multline}
%\label{eq:qffinvalid}
%[y < 0] \lor [w > 0] \lor [x > 0] \lor [z < 0] 
% \\ 
%\lor [ x + 1 < 0 \land y - 1 > 0 ] \lor [ w + 1 < 0 \land z - 1 > 0 ] 
% \\ 
%\lor [ w + 1 < 0 \land y w - w + y + x \geq  0 \land x z + z - y w + w - y - x > 0 ] 
% \\ 
%\lor  [ y w - w + y + x < 0 \land z - 1 > 0 \land x z + z - y w + w - y - x < 0 ] 
%\\ 
%\lor [ y - 1 > 0 \land y w - w + y + x < 0 ].
%\end{multline}
\begin{align}
&[y < 0] \lor [w > 0] \lor [x > 0] \lor [z < 0] 
\nonumber \\ 
&\lor [ x + 1 < 0 \land y - 1 > 0 ] \lor [ w + 1 < 0 \land z - 1 > 0 ] 
\nonumber \\ 
&\lor [ w + 1 < 0 \land y w - w + y + x \geq  0 
\nonumber \\ &\qquad \land x z + z - y w + w - y - x > 0 ] 
\nonumber \\ 
&\lor  [ y w - w + y + x < 0 \land z - 1 > 0 
\nonumber \\ &\qquad \land x z + z - y w + w - y - x < 0 ] 
\nonumber \\ 
&\lor [ y - 1 > 0 \land y w - w + y + x < 0 ].
\label{eq:qffinvalid}
\end{align}

%Note that we can use {\sc Qepcad} to eliminate the spurious variable $t$ from just the final (quantified) formula in \eqref{eq:nonprenexinvalid} (just the formula describing case D). This takes 1063 cells and 0.2 seconds to calculate and returns an equivalent answer to \eqref{eq:qffinvalid}. This is quicker but produces more cells than \eqref{eq:prenexinvalid} as it cannot take full advantage of {\sc Qepcad}'s in-built tools (such as formula simplification).

\subsection{New formulation for CAD}

%subsection{Negation of invalid regions}

We now have a description of the invalid regions, \eqref{eq:qffinvalid}, so we can describe the valid regions by taking its negation:
%\begin{multline}\label{eq:negatedinvalid}
%[w\leq 0] \land [x\leq 0] \land [y\geq 0] \land [z\geq 0] 
% \land [x\geq -1 \lor y\leq 1] \\ \land [w\geq -1 \lor z\leq 1] 
% \land \big[w y - w + x + y < 0 \lor w+1 \geq 0 \\ \lor xz+z-yw+w-y-x \leq 0 \big] 
% \land \Big[ yw-w+y+x \geq 0 \\ \lor \big[ [z-1\leq 0 \lor xz+z-yw+w-y-x \geq 0] \land y-1\leq 0   \big]   \Big]
%\end{multline}
\begin{align}
&\quad [w\leq 0] \land [x\leq 0] \land [y\geq 0] 
\land [z\geq 0] \land [x\geq -1 \lor y\leq 1] 
\nonumber \\ 
&\land [w\geq -1 \lor z\leq 1] 
\land \big[w y - w + x + y < 0 \lor w+1 \geq 0 
\nonumber \\
&\quad \lor xz+z-yw+w-y-x \leq 0 \big] 
\nonumber \\
&\land \big[ yw-w+y+x \geq 0 
\lor \big[ [z-1\leq 0 \nonumber \\
&\quad \quad \lor xz+z-yw+w-y-x \geq 0] \land y-1\leq 0   \big]   \big].
\label{eq:negatedinvalid}
\end{align}

%subsection{New formulation}

Although (\ref{eq:negatedinvalid}) describes the valid regions in terms of the endpoints it is missing any description of the relationship between these (fixing the length of the ladder).  Hence our new formulation of the problem for CAD is 
\begin{equation}
\label{eq:newformulation}
[(x - w)^2 + (y - z)^2 = 9] \land \eqref{eq:negatedinvalid}.
\end{equation}

%\begin{multline}\label{eq:newformulation}
%[(x - w)^2 + (y - z)^2 = 9] \land
%[w\leq 0] \land [x\leq 0] \land [y\geq 0] \land [z\geq 0] \\
% \land [x\geq -1 \lor y\leq 1] \land [w\geq -1 \lor z\leq 1] 
% \land \big[w y - w + x + y < 0 \\ \lor w+1 \geq 0 \lor xz+z-yw+w-y-x \leq 0 \big] 
% \land \Big[ yw-w+y+x \geq 0 \\ \lor \big[ [z-1\leq 0 \lor xz+z-yw+w-y-x \geq 0] \land %y-1\leq 0   \big]   \Big] 
%\end{multline}
%(which is simply $[(x - w)^2 + (y - z)^2 = 9] \land \eqref{eq:negatedinvalid}$).

\subsection{Applying CAD}

The formula \eqref{eq:newformulation} was given to {\sc Qepcad} (with initialisation parameters 
{\tt +N500000000 +L200000}) under the variable ordering $x \prec y \prec w \prec z$. After a little under 5 hours (16,933.701 seconds) of computation time a CAD of $\R^4$ was constructed with 285,419 cells.  The following equivalent formula to \eqref{eq:newformulation} was given:
%\begin{multline}\label{eq:solution}
%x \leq 0 \land y \geq 0 \land w \leq 0 \land z \geq 0 \land (y-z)^2+(x-w)^2 = 9 \\  
%\land \Big[ [ x + 1 \geq 0 \land w + 1 \geq 0 ] \lor \big[ y - 1 \leq 0 \land w + 1 \geq 0 \\ 
%\land y^2 w^2 - 2 y w^2 + x^2 w^2 + 2 x w^2 + 2 w^2 - 2 x y^2 w + 4 x y w - 2 x^3 w \\ - 4 x^2 w 
%- 4 x w + x^2 y^2 - 2 x^2 y + x^4 + 2 x^3 - 7 x^2 - 18 x - 9 \geq 0 \big] \lor \\ \big[ x + 1 \geq 0 
%\land y w - w + y + x \geq 0 \land w^2 - 2 x w + y^2 - 2 y + x^2 - 8 > 0 \\ \land z - 1 \leq 0 \big] 
%\lor \big[ x + 1 \geq 0 \land y w - w + y + x \geq 0 \land y^2 w^2 - 2 y w^2 \\+ x^2 w^2 + 2 x w^2 + 2 w^2
%- 2 x y^2 w 
%+ 4 x y w - 2 x^3 w - 4 x^2 w - 4 x w \\+ x^2 y^2 - 2 x^2 y + x^4 + 2 x^3 - 7 x^2 - 18 x - 9 \leq 0 
%\land z - 1 \leq 0 \big]\\ \lor [ y - 1 \leq 0 \land z - 1 \leq 0 ] \Big]
%\end{multline}
\begin{align}
&\,\, x \leq 0 \land y \geq 0 \land w \leq 0 \land z \geq 0 \land (y-z)^2+(x-w)^2 = 9 
\nonumber \\
&\land \Big[ [ x + 1 \geq 0 \land w + 1 \geq 0 ] 
\lor \big[ y - 1 \leq 0 \land w + 1 \geq 0 
\nonumber \\ 
&\quad \quad \land 
y^2 w^2 - 2 y w^2 + x^2 w^2 + 2 x w^2 + 2 w^2 - 2 x y^2 w 
\nonumber \\ 
&\quad \quad \quad + 4 x y w - 2 x^3 w - 4 x^2 w - 4 x w + x^2 y^2 - 2 x^2 y 
\nonumber \\
&\quad \quad \quad + x^4 + 2 x^3 - 7 x^2 - 18 x - 9 \geq 0 \big] 
\nonumber \\ 
&\quad \lor \big[ x + 1 \geq 0 
\land y w - w + y + x \geq 0 \land w^2 - 2 x w + y^2 
\nonumber \\ 
&\quad \quad - 2 y + x^2 - 8 > 0 \land z - 1 \leq 0 \big] 
\nonumber \\
&\quad \lor \big[ x + 1 \geq 0 \land y w - w + y + x \geq 0 \land y^2 w^2 - 2 y w^2 
\nonumber \\
&\quad \quad + x^2 w^2 + 2 x w^2 + 2 w^2 - 2 x y^2 w + 4 x y w - 2 x^3 w 
\nonumber \\
& \quad \quad \quad - 4 x^2 w - 4 x w + x^2 y^2 - 2 x^2 y + x^4 + 2 x^3 - 7 x^2 
\nonumber \\ 
& \quad \quad \quad - 18 x - 9 \leq 0 \land z - 1 \leq 0 \big]
\nonumber \\ 
&\quad \lor [ y - 1 \leq 0 \land z - 1 \leq 0 ] \Big].
\label{eq:solution}
\end{align}

The first line gives the conditions of the problem which are in conjunction with any valid configuration.  The remaining lines give a large disjunction of clauses describing such configurations.  The first clause is characterizing the positions where the ladder is entirely in the vertical corridor and the last clause where the ladder is entirely in the horizontal corridor. There are then three more clauses characterising positions in between. Any analysis of the decomposition of these equations requires knowledge of the adjacency of the four-dimensional CAD: this is highly non-trivial and discussed in Section \ref{subsec:adjacency}.

{\sc Qepcad} uses, amongst other theory, partial CAD techniques %\cite{Collins:1991vz}
\cite{CH91} %and equational constraints (\cite{McCallum:1999hj}) 
to simplify its calculations and output. These can be %(at least partially) 
suppressed by issuing the {\tt full-cad} command. We note that doing so greatly increases the difficulty of the problem.
Calculating a {\tt full-cad} of \eqref{eq:newformulation} resulted in the construction of 1,691,473 cells taking just over a day of computation time (88,238.442 seconds). The quantifier-free formula returned is almost identical to the partial CAD version \eqref{eq:solution} (with a couple of cases split slightly differently). 

We can attempt to speed up the construction by introducing quantifiers on  one endpoint leading to a CAD of valid positions for one endpoint of the ladder, by prefixing \eqref{eq:newformulation} with $(\exists w)(\exists z)$. 
% (This approach was taken for the problem in \eqref{eq:originalpiano} by [XYZ].)  
Using {\sc Qepcad} this took just over 50 minutes (3052.753 seconds) and produced only 5453 cells.  The sharp reduction is  a result of partial CAD techniques as described in %\cite{Collins:1991vz}
\cite{CH91}. The resulting quantifier-free formula is simply,
\begin{equation}\label{eq:singleendpoint}
x \leq 0 \land y \geq 0 \land [ x + 1 \geq 0 \lor y - 1 \leq 0 ],
\end{equation}
which is the definition of the original corridor.  The quantified version of \eqref{eq:newformulation} is simply asking for those points where it is possible to place an end of the ladder and have it in a valid position and so this formula is as expected.  We note that the CAD used to construct the formula contains far more information than is needed --- a CAD with only 17 cells is sufficient to describe the corridor.   

The existential CAD is not sufficient to solve the path finding problem, and for our example the output \eqref{eq:singleendpoint} gives little useful information. However, providing quantified variables has drastically reduced the complexity of the problem and so can be a useful test for the feasibility of the problem (a CAD for the original formulation \eqref{eq:originalpiano} remains infeasible under quantification). It can also be used in some cases (when the valid region for the endpoint is not the entire corridor) to show a ladder traversal is impossible: for example, if an invalid region were to 'block' the corridor.

\begin{figure}
\centering
\includegraphics[scale=0.7]{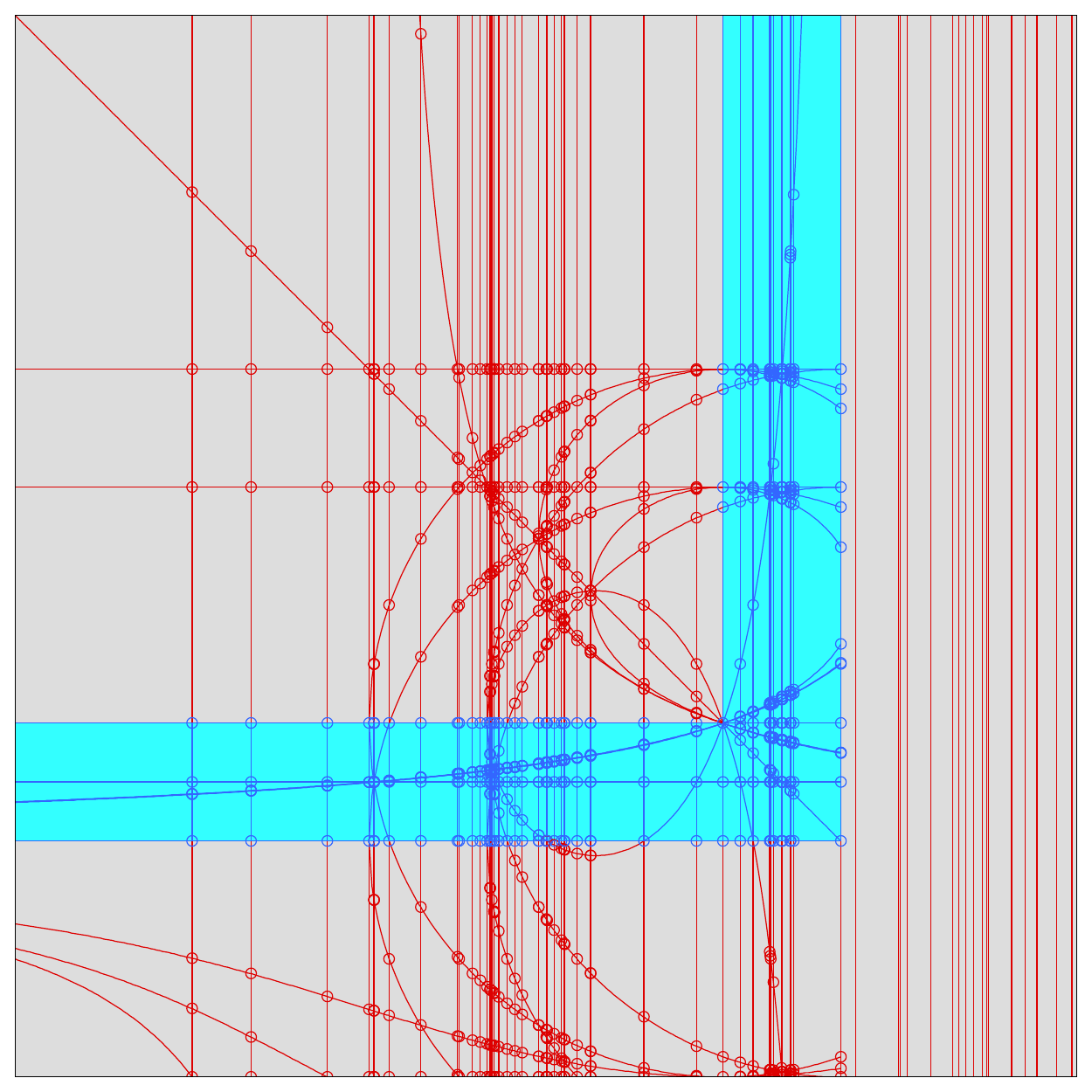}
\caption{A two-dimensional CAD of the $(x,y)$ configuration space constructed from \eqref{eq:newformulation}.}\label{fig:existentialcad}
\end{figure}

{\sc Qepcad} can produce a visualisation of two-dimensional CADs through the {\tt p-2d-cad} command. Figure \ref{fig:existentialcad} shows the output for the problem in the preceding paragraphs (so it refers to the existential CAD; the diagram for the non-quantified formulation is similar, but omits all cell boundaries within the corridor). The diagram is for $x$ in the range $[-7,2]$ and $y$ in the range $[-2,7]$ with a step of $0.025$ (therefore if stacks are within $0.025$ (with respect to $x$) or intra-stack cells are within $0.025$ (with respect to $y$) they will not be distinguishable). 

Figure \ref{fig:existentialcad} makes clear just how complicated the problem is when being tackled by CAD. There are certainly boundaries to cells that seem to be related to `boundary cases' of the problem: when the ladder is `stuck' trying to get around the corner.  However, there are many boundaries with little significance for the real problem and so further development of the CAD technology to remove these would be beneficial.

\subsection{Adjacency}
\label{subsec:adjacency}

The four-dimensional CADs of configuration space we have produced from \eqref{eq:newformulation} could be used to both determine the existence of a solution and then construct a path.  However, to do the latter we need to first analyse the adjacency and connectedness of cells in the four-dimensional CAD.  This is not currently possible with any existing technology and is certainly non-trivial. The process is described in two dimensions by \cite{ACM84II} (which has been implemented in {\sc Qepcad}) while \cite{ACM88} generalises the approach to three dimensions.  Further generalisations are not trivial however \cite{BGV13}, with adjacency algorithms likely to work (without a change of coordinates) only for well-behaved input. We also note that in \cite{SS83II} the authors consider adjacencies between $n$ and $(n-1)$-dimensional cells, but since we have an equational constraint, we are actually interested in adjacencies between $(n-1)$ and $(n-2)$-dimensional cells.

\subsection{Choosing a formulation}
\label{subsec:heuristics}

An important question is why \eqref{eq:newformulation} is a better formulation for CAD than \eqref{eq:originalpiano}, and whether we could have predicted this.

On first glance, we see that the new formulation involves polynomials of lesser degree. One measure of CAD complexity is {\tt sotd} (introduced in %\cite{Dolzmann:2004fc}
\cite{DSS04} as the sum of total degree of each monomial in each polynomial).  Using this measure applied to the input polynomials as a heuristic certainly favours the new formulation: \eqref{eq:originalpiano} has {\tt sotd} 100 compared to \eqref{eq:newformulation} with an {\tt sotd} of 33.  The benefit is less obvious when taking an {\tt sotd} of the full projection factor sets. The new formulation is still lower, but there is a smaller relative difference: 2006 is reduced to 1693.
There are over 100 univariate polynomials in the projection sets of both formulations. Calculating {\tt ndrr} (introduced in %\cite{Bradford:2013bb}
\cite{BDEW13} as the number of distinct real roots of the univariate projection polynomials) also favours the new formulation, but again, not by an amount that indicates the changes in feasibility: 367 reduces to 301.

% which considers formulation of CAD problems in a more general setting) directly is costly. If we calculate the {\tt ndrr} of each polynomial separately (possibly counting roots repeatedly) we get a relatively small difference that still indicates the new formulation as better: 

For comparison, we note that the approach by Wang leads to an {\tt sotd} of 19 for the the top level projection polynomials, 98 for the full projection factor set and an {\tt ndrr} of 17. McCallum's formulation has {\tt sotd}'s of 68 and 32 (lower due to repeated factors) and an {\tt ndrr} of 5. Yang-Zeng's approach gives {\tt sotd}'s of 35 and 39, and an {\tt ndrr} of 2. Hence full {\tt sotd} and {\tt ndrr} correctly predict that the CADs related to these approaches will be smaller than our reformulation.

These heuristics do not take into account the number of quantifiers which can be hugely influential in the complexity of a problem. The fact that Wang's formulation contained only a single unquantified variable is hugely instrumental in such an efficient construction.
The effect of these quantifiers suggests the creation of more sophisticated heuristics. For example: sum of weighted total degrees. This would weight variables according to two properties: the overall variable ordering and which variables are quantified. 

Let the CAD be created with respect to variables $x_1 \prec x_2 \prec \cdots \prec x_n$ where $x_1$ decomposes $\mathbb{R}^1$, $\{x_1,x_2\}$ decomposes $\mathbb{R}^2$ and so forth. Then assign a weight of $i$ to variable $x_i$ so that the polynomial $x_5^3 - x_1$ would have {\tt sowtd} 16 rather than just an {\tt sotd} of 4. In addition to this, if a variable is quantified then reflect this by halving its effect on {\tt sowtd}. For the above polynomial, if $x_5$ was quantified then the ${\tt sowtd}$ would become 8.5.
Applying these to the various formulations we get the following:
\begin{itemize}
  \item Davenport (unquantified): ${\tt sowtd} = 148$.
  \item Davenport (quantified): ${\tt sowtd} =92$.
  \item New formulation (unquantified): ${\tt sowtd}=72$.
  \item McCallum's formulation: ${\tt sowtd}=70$.
  \item New formulation (quantified): ${\tt sowtd} = 46$.
  \item Wang's formulation: ${\tt sowtd} = 27$.
  \item Yang--Zeng's formulation: ${\tt sowtd} = 23$.
\end{itemize}
The {\tt sowtd} measure gives an ordering matching the difference in cell counts, and has plausible-looking differences.
\section{Generalising the problem}
\label{sec:Generalise}

\subsection{Ladders of different length}
\label{subsec:length}

The reformulation described in Section \ref{sec:NewForm} was for a ladder of length $3$. We know already that the maximum length of a ladder able to traverse the corner is $\sqrt{8}$ and similar geometric reasoning shows that the maximum length of a ladder able to reverse its orientation is $\sqrt{2}$.  We compare the CAD for \eqref{eq:newformulation} (in which the ladder can not traverse the corridor) to the equivalent formulations with a ladder of shorter length.  We consider four canonical cases which exhaust the possible scenarios:
\begin{description}
\item[Length 3:] \quad Ladder cannot traverse the corridor.
\item[Length 2:] \quad Ladder can traverse the corridor but is unable to reverse its orientation.
\item[Length $\frac{5}{4}$:] \quad Ladder can traverse the corridor and is able to reverse its orientation, but only within the `corner'.
\item[Length $\frac{3}{4}$:] \quad Ladder can traverse the corridor and reverse its orientation at any point within the corridor.
\end{description}

All the results are summarized in Table \ref{tab:results}.  We compare both non-quantified and quantified versions (where the input formula was preceded by $(\exists w)(\exists z)$ as indicated by $\exists$).
Note the length of the ladder is an explicit equational constraint and so {\sc Qepcad} automatically applies the theory of \cite{McCallum99}.
%, but an equational constraint can be explicitly designated. The number of cells does not change when the equational constraint is declared explicitly, but there is a slight speed up in time.
%\begin{table}[b]
%\centering
%\caption{Results for solving \eqref{eq:newformulation} with varying lengths. EC indicates that the equational constraint was explicitly stated.}\label{tab:results1}
%\begin{tabular}{|c|cc|cc|}
%\hline & \multicolumn{2}{c}{CAD} & \multicolumn{2}{|c|}{EC-CAD} \\
%  Length & Cells & Time (s) & Cells & Time (s) \\\hline 
% 3 & 285419 & 16933.701 & 285419 & 16286.431 \\
% 2 & 314541 & 10231.070 & 314541 & 9863.950 \\
% 5/4 & 404449 & 34288.130 & 404449 & 33042.101\\
% 3/4 & 446787 & 13652.885 & 446787 & 13146.195 \\\hline
% 
% 3 {\tt full-cad} & 1691473 & 88238.442 & --- & --- \\\hline
%\end{tabular}
%\end{table}
%\begin{table}[b]
%\centering
%\caption{Results for solving the existential version (input formula was preceded by $(\exists w)(\exists z)$) of \eqref{eq:newformulation} with varying lengths. EC indicates that the equational constraint was explicitly stated .}\label{tab:results2}
%\begin{tabular}{|c|cc|cc|}
%\hline
% & \multicolumn{2}{|c}{$\exists$ CAD} & \multicolumn{2}{|c|}{$\exists$ EC-CAD}\\
%  Length & Cells & Time (s) & Cells & Time (s) \\\hline 
% 3 & 5453 & 3052.753 & 5453 & 2941.024 \\
% 2 & 5353 & 1997.280 & 5353 & 1922.837 \\
% 5/4 & 5589 & 7559.598 & 5589 & 7312.347 \\
% 3/4 & 4347 & 72.282 & 4347 & 69.690 \\\hline
%
%\end{tabular}
%\end{table}

\begin{table}[h]
\centering
\caption{
CADs of \eqref{eq:newformulation} modified by varying ladder length.  
%We compare both non-quantified and quantified versions (where the input formula was preceded by $(\exists w)(\exists z)$ as indicated by $\exists$ in the table).
}
\label{tab:results}
\begin{tabular}{|c|rr|rr|}
\hline  & \multicolumn{2}{|c|}{EC-CAD} & \multicolumn{2}{|c|}{$\exists$ EC-CAD} \\
  Length & Cells & Time (s) & Cells & Time (s) \\\hline 
 3 & 285419 & 16286.431 & 5453 & 2941.024 \\
 2 & 314541 & 9863.950 & 5353 & 1922.837 \\
 5/4 & 404449 & 33042.101 & 5589 & 7312.347 \\
 3/4 & 446787 & 13146.195 & 4347 & 69.690 \\\hline
 
 3 {\tt full-cad} & 1691473 & 88238.442 & --- & --- \\\hline
\end{tabular}
\end{table}

\subsection{Angled corridors}
\label{subsec:angled}

%subsubsection{General Angled Corridor}

\begin{figure}[b]
\centering
\begin{tikzpicture}
\draw[ultra thick] (0,0)--(-2,0);
\draw[ultra thick] (0,0)--(2,1);
\draw[ultra thick] (0,1)--(-2,1);
\draw[ultra thick] (0,1)--(2,2);

\draw[dashed] (0,0) -- (1,0);
\draw[dashed] (0,1) -- (1,1);

\draw (0.5,0) arc(0:25:0.5);
\draw (0.5,1) arc(0:25:0.5);

\node [right] at (0.5,0.15) {{\small $\theta$}};
\node [right] at (0.5,1.15) {{\small $\theta$}};

\end{tikzpicture}
\caption{Generic obtuse angled corridor}
\label{fig:obtuseangle}
\end{figure}
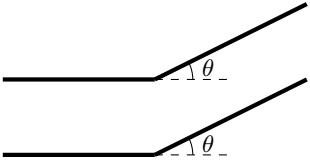

\begin{figure}[b]
\centering
\begin{tikzpicture}
\draw[ultra thick] (-0.25,0)--(-4,0);
\draw[ultra thick] (-0.25,0)--(-2.25,4);
\draw[ultra thick] (-2,1)--(-4,1);
\draw[ultra thick] (-2,1)--(-3.5,4);

\draw (-0.75,0) arc(180:115:0.5);
\draw (-2.5,1) arc(180:115:0.5);

\node [left] at (-0.25,0.15) {{\small $\psi$}};
\node [left] at (-2,1.15) {{\small $\psi$}};
\end{tikzpicture}
\caption{Generic acute angled corridor}
\label{fig:acuteangle}
\end{figure}
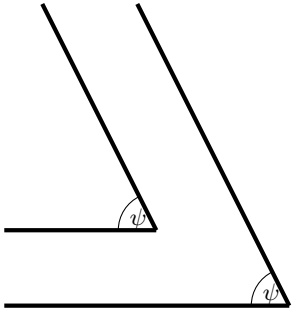

We consider how the problem may be generalised to a non-right angled corridor.  There are two canonical cases: that where the angle is obtuse as in Figure \ref{fig:obtuseangle} and that where the angle is acute as in Figure \ref{fig:acuteangle}.  

For a general obtuse angled corridor the right hand corridor walls have equations $y = \tan(\theta)x$ and $y = \tan(\theta)x + 1$. This results in the following formulation of the invalid positions:
%\begin{multline}
%[ x < 0 \land y > 1] \lor [y < 0] \lor [x > 0 \land y > \tan(\theta) x + 1] \lor \\ 
%[y < \tan(\theta)x] \lor 
%[ w < 0 \land z > 1] \lor [z < 0] \lor \\
%[w > 0 \land z > \tan(\theta) w + 1] \lor [z < \tan(\theta)w] \lor \\
%(\exists t) [0 < t \land t < 1] \land \Big[ [x+t(w-x) < 0 \land y+t(z-y)> 1] \\
%\lor [y+t(z-y) < 0] \lor  [x+t(w-x) > 0 \land y+t(z-y) > \\ \tan(\theta)(x+t(w-x))+1] \lor  [y+t(z-y) < \tan(\theta)(x+t(w-x))   \Big]
%\end{multline}

\begin{align}
&\quad [ x < 0 \land y > 1] \lor [y < 0] \lor [x > 0 \land y > \tan(\theta) x + 1] 
\nonumber \\ 
&\lor [y < \tan(\theta)x] \lor [ w < 0 \land z > 1] \lor [z < 0] 
\nonumber \\
&\lor [w > 0 \land z > \tan(\theta) w + 1] \lor [z < \tan(\theta)w]
\nonumber \\
&\lor (\exists t) [0 < t \land t < 1]
\nonumber \\
&\quad \land \Big[ [x+t(w-x) < 0 \land y+t(z-y)> 1]
\nonumber \\
&\quad \quad \lor [y+t(z-y) < 0] \lor  [x+t(w-x) > 0
\nonumber \\
&\quad \quad \quad \land y+t(z-y) > \tan(\theta)(x+t(w-x))+1]
\nonumber \\
&\quad \quad \lor  [y+t(z-y) < \tan(\theta)(x+t(w-x))]   \Big].
\end{align}

%Whilst it may be infeasible to construct such a CAD, it is simple to construct a formulation.

For a general acute angled corridor the right hand corridor walls have equations $y = -\tan(\psi)x$ and $y = -\tan(\psi)(x + 1)$. This results in the following formulation of the invalid positions:
%\begin{multline}
%[y < 0] \lor [y > -\tan(\psi)x] \lor \big[ x < -\left( \frac{\tan(\psi) + 1}{\tan(\psi)} \right)  \land \\
% y > 1 \land y < -\tan(\psi) (x+1) \big] \lor [z < 0] \lor [z > -\tan(\psi)w] 
%\\
%\big[ w < -\left( \frac{\tan(\psi) + 1}{\tan(\psi)} \right)  \land 
%z > 1 \land z < -\tan(\psi) (w+1) \big] \lor \\
%(\exists t) [0 < t \land t < 1] \land \Big[  x + t(w-x) <  -\left( \frac{\tan(\psi) + 1}{\tan(\psi)} \right) \land \\
%y +t(z-y) > 1 \land y+t(z-y) < -\tan(\psi)(x+t(w-x)+1) \Big]
%\end{multline}
\begin{align}
&\quad [y < 0] \lor [y > -\tan(\psi)x]
\nonumber \\
&\lor \left[ x < -\left( \frac{\tan(\psi) + 1}{\tan(\psi)} \right) 
\land y > 1 \land y < -\tan(\psi) (x+1) \right] 
\nonumber \\
&\lor [z < 0] \lor [z > -\tan(\psi)w] 
\nonumber \\
&\lor \left[ w < -\left( \frac{\tan(\psi) + 1}{\tan(\psi)} \right)  \land 
z > 1 \land z < -\tan(\psi) (w+1) \right] 
\nonumber \\
&\lor (\exists t) [0 < t \land t < 1] 
\land \Big[  x + t(w-x) <  -\left( \frac{\tan(\psi) + 1}{\tan(\psi)} \right) 
\nonumber \\
&\quad \land y +t(z-y) > 1 
\nonumber \\
&\quad \land y+t(z-y) < -\tan(\psi)(x+t(w-x)+1) \Big].
\end{align}

If $\tan$ of the angle in question is an algebraic number (for example if the angle is a rational multiple of $\pi$) then we can compute an exact solution to these problems using CAD. However for other cases we would either need to approximate the value of $\tan(\theta)$ or treat it as an additional variable in configuration space.  

As with the formulation for the right angled corridor we then eliminate the extra parameter $t$, take the negation of the quantifier free formula, conjunct the equational constraint describing the length of the ladder, and construct a CAD according to this new formula.  Hence the new formulation in Section \ref{sec:NewForm} may be generalised easily, although constructing the CAD may be more computationally difficult. Generalising Wang and Yang--Zeng's methods is not always straightforward due to them being so reliant on geometrical reasoning, as demonstrated in the examples below. It should be possible to adapt \cite{McCallum97} for angled corridors, although care may need to be taken that certain trigonometric identities hold.

\subsubsection{Obtuse $\pi/4$-angled corridor}

Let the walls of the right angled corridor make an angle of $\pi/4$ with the horizontal.

We can generalise Wang's idea and consider when the ladder intersects all four walls at once. As with the right-angled corridor, this provides us with the maximal length of the ladder. This approach would work for all obtusely angled corridors (under the same constraint of $\tan(\theta)$ being algebraic).

{\sc Qepcad} can answer this question with 27 cells in 5.717 seconds to return
\[
r = 0 \lor 2r^6 - 93r^4 - 172r^2 - 125 \geq 0.
\]

The appropriate solution is
\[
\sqrt{ \frac{1}{6} \sqrt[3]{667143 + 4452 \sqrt{159}} + \frac{(2539/2)}{\sqrt[3]{667143+4452\sqrt{159}}} + \frac{31}{2}}
\]
which is approximately 6.6786.

Tackling this problem with our method, we first we eliminate $t$ from the invalid regions. This takes 170,597 cells and 230.881 seconds. After forming the complete formulation, {\sc Qepcad} fails to construct the relevant CAD after constructing 50,000,000 cells (the self-imposed limit of the {\tt +N50000000} parameter when calling {\sc Qepcad}). The extra complexity is because the diagonal corridor is not aligned with the directions of projection.

\subsubsection{Acute $\pi/4$-angled corridor}

Let the walls of the right angled corridor make an angle of $3\pi/4$ with the horizontal to form an acutely angled corridor with angle $\pi/4$.

We can na\"ively apply Wang's method to the acutely angled corridor. {\sc Qepcad} uses 39 cells in 4.520 seconds to return
\[
r=0 \lor 2r^6 + 9r^4 -17r^2 -125 \geq 0.
\]
The appropriate solution is 
\[
\sqrt{ \frac{1}{6} \sqrt[3]{4644 + 249 \sqrt{249}} + \frac{61}{2 \sqrt[3]{4644+249\sqrt{249}}} - \frac{3}{2}}
\]
which is approximately 1.8443.

Unfortunately this does not give a complete answer to the problem.  It is possible to fit a rod of length $\sqrt{5}$ (greater than the above value) by placing it within the corner. This disparity is because na\"ively applying Wang's idea does not take into account the possibility of reversing the orientation of the ladder necessary for ladders of larger lengths. To adapt Wang's idea to include this reverse orientation would require some non-trivial geometric reasoning. 

As our formulation is based within configuration space, it acknowledges the extra condition of orientation, but with added complexity expense. If we try our formulation with $r=2$ (as $1.8443 < 2 < \sqrt{5}$) we first eliminate $t$ from the invalid regions. This takes 91,583 cells and 86.647 seconds. If we then try to solve the problem by constructing the relevant CAD we fail after constructing 50,000,000 cells (the self-imposed limit of the {\tt +N50000000} parameter when calling {\sc Qepcad}).

\section{Adapting CAD technology for future Work}
\label{sec:NewTech}

{\sc Qepcad} makes use of the theory of equational constraints to reduce the number of projection polynomials (and hence the number of cells in the CAD) along with partial CAD techniques.  However, we note that there are further savings that could be made given the presence of the equation and we discuss these ideas and their potential in this section.   

\subsection{Extending Equational Constraints}

First, as pointed out in \cite{England13b}, the theory of \cite{McCallum99} allows us to only lift with respect to the equational constraint for the the final lift.  However, \textsc{Qepcad} appears to lift with respect to all projection polynomials (including the non-equational constraints).  Considerable savings can be made by implementing this idea. If more than one equational constraint is present in a problem (for example if there were multiple ladders) then the full power of TTICAD (as described in \cite{BDEMW13}) can be used to simplify the resulting CAD further.

\subsection{Building a layered CAD for the problem}

As mentioned earlier we are concerned with the adjacencies and connectedness of our CAD of configuration space. 
For this problem we are mainly concerned with those cells in the CAD of full-dimension as these describe regions where the configuration of the ladder is free to move.  We note that the key adjacencies for these cells are those through two-dimensional cells: an adjacency of two three-dimensional cells through a one- or zero-dimensional cell would correspond to an infeasible situation in the real physical space for all but boundary cases (i.e. the ladder having to ``tightrope walk'' a one-dimensional subspace of $\mathbb{R}^2$). 

The idea of building CADs containing only cells of full-dimension has been investigated previously in \cite{McCallum93, Strzebonski00, Brown13}.  We have generalised the idea to produce CADs with cells of specified dimension and higher, which we call \emph{layered CADs}.  Algorithms to produce these are presented in \cite{WE13} along with a discussion of their topological properties, possible applications and an implementation in {\sc Maple} built over the authors' {\tt ProjectionCAD} package, \cite{England13a, England13b}.  Work on these objects and their properties is ongoing.

\subsection{Lifting to a manifold}

In the configuration space, all valid cells must lie on the three-dimensional manifold described by the equation $(x-w)^2+(y-z)^2=9$ so we are only concerned with cells where the equation is satisfied (and the ladder has the desired length). 
We can therefore construct an order-invariant CAD of three-dimensional space using the projection polynomials for input with an equational constraint \cite{McCallum99}, and when lifting over this with respect to the equation, discard all sectors. This leaves just the sections: precisely the cells on the manifold.  We have implemented this approach using our \textsc{Maple} package \cite{England13b}.

Within the manifold, the most important cells are those of full-dimension (with respect to the manifold) as cells of a lower dimension relate to physically infeasible situations (i.e. one-dimensional subspaces of $\mathbb{R}^2$). We can restrict our CAD to produce only these cells through a smarter lifting stage. 
%We show this method with our new formulation.

Any full-dimensional cell on the three-dimensional manifold must project onto a three-dimensional cell in the induced CAD of $\mathbb{R}^3$ (as it is a section of the equational constraint). We start by constructing the projection set with respect to the equational constraint, producing 11 polynomials in $y,w,z$. We then build just the full-dimensional CAD cells in $\mathbb{R}^3$: 64,764 cells in 16,991.400 seconds. 

We can now lift over these cells with respect to the manifold (an equational constraint). We construct a stack over each cell and extract any sections (those cells lying on the manifold). This process is relatively quick, produces 101,924 cells in 1020.860 seconds. The total time to construct the three-dimensional decomposition of the manifold is therefore 18,012.3 seconds, producing 101,924 cells. 

It is not yet feasible to construct all cells on the manifold (or indeed the three-dimensional CAD to lift over) using {\sc Maple}, partly as our implementation does not yet take advantage of partial CAD techniques.  We expect that with further improvements a CAD of the manifold sufficient for constructing valid paths (one with two and three-dimensional cells) could be built.

\section{Conclusions}
\label{sec:Conclusion}

We considered a classic example of a piano mover's problem, how a ladder can traverse a corridor, and the solution via CAD.  Despite years of improvements to CAD theory and computer hardware, building a CAD for the original formulation in \cite{Davenport86} remains infeasible.  However, by reformulating the problem CADs can be produced in a matter of seconds, demonstrating how problem formulation is essential to the feasibility of a CAD problem.  Further evidence of this was presented in \cite{BDEW13}.

We presented a new formulation of the problem for which a CAD can be produced.  There are other solutions in the literature \cite{Wang96, McCallum97, YZ06} but these differ in important ways.  In \cite{Wang96, YZ06} the authors relied heavily upon mathematical deduction performed by hand before input to CAD.  While this is the most powerful reformulation tool available it is not trivial to automate or generalise. In \cite{McCallum97} the author described configurations in a non-trivial manner involving translations and rotations which may complicate subsequent analysis of the space.  Our reformulation in Section \ref{sec:NewForm} uses only a simple negation of the problem, a technique that could be performed algorithmically by CAD technology, using heuristics to decide the appropriate formulation to use.  

Another distinction is that the approaches in \cite{Wang96, YZ06} are firmly rooted within the two-dimensional space of the corridor while the new formulation presented here deals with the configuration space of the ladder: a three-dimensional manifold within four-dimensional space.  This means that whilst the approaches in \cite{Wang96, YZ06} are able to answer the question ``Can the ladder get through the corridor?'' they cannot answer the question ``How can the ladder get through the corridor?''.  The approach of \cite{McCallum97} would be able to answer the latter question, but only after some non-intuitive analysis of the trigonometric space described by the formulation.

These distinctions may seem trivial for the problem at hand, but they would become far more important in generalisation.  Indeed, even for a ladder in an acute-angled corridor it may be that the only feasibility path involves rotating the ladder in the corner (reversing its orientation).  This would not be provided by a simple affirmation that a path existed and in the case of Wang's formulation the possibility would not be considered since this formulation requires the orientation to be fixed.  It is hard to think of a mathematical argument that takes this into account without needing the full configuration space. 

Finally, we have also introduced the idea of restricting lifting in CAD, to cells of full dimension lying on the given manifold.  This is much more efficient than producing a full CAD, returning just over a third of the cells for our example. These techniques could be applied to any problem with an equational constraint (lifting to appropriate manifold, or indeed hypersurface) and have now been investigated in generality and formalised in \cite{WBDE14}.

Although a generic symbolic solution to robot motion planning was provided in theory by \cite{SS83II}, in general it remains infeasible to the present day, with numerical methods providing the only practical approach.  The ideas presented in this paper show that progress is still possible, but that it will likely follow from more appropriate formulations of problems just as much as advances in theory and technology.  

\section*{Acknowledgements}
This work was supported by EPSRC grant: EP/J003247/1.

%\bibliographystyle{amsalpha}
%\bibliography{../../DJW}
%\bibliography{DJW}
\bibliographystyle{plain}
\bibliography{CAD}

\end{document}